# Persistent incommensurate amorphous/crystalline meta-interfaces enable engineering-grade superlubricity

*Wan Wang, Zijun Ding, Panpan Li,\* Wanying Ying, Hongxuan Li, Xiaohong Liu, Huidi Zhou, Jianmin Chen, Wengen Ouyang,\* and Li Ji\**

W. Wang, P. Li, H. Li, X. Liu, H. Zhou, J. Chen, L. Ji
State Key Laboratory of Solid Lubrication, Lanzhou Institute of Chemical Physics, Chinese Academy of Sciences, Lanzhou 730000, P. R. China
E-mails: lipanpan@licp.cas.cn; jili@licp.cas.cn.

Z. Ding, W. Ying, W. Ouyang
Department of Engineering Mechanics, School of Civil Engineering, Wuhan University, Wuhan, Hubei, 430072, P. R. China
E-mail: w.g.ouyang@whu.edu.cn

W. Ouyang
State Key Laboratory of Water Resources Engineering and Management, Wuhan University, Wuhan, Hubei, 430072, P. R. China

W. Wang, H. Li, X. Liu, L. Ji
Center of Materials Science and Optoelectronics Engineering, University of Chinese Academy of Sciences, Beijing 100049, P. R. China
W. Wang and Z. Ding contributed equally to this work

**Keywords:** superlubricity, meta-contacts, extreme conditions, amorphous/crystalline

**Abstract:** Friction dissipates a substantial portion of global energy, motivating the pursuit of superlubricity, a state of near-zero friction, in real-world systems. Conventional approaches rely on crystalline lattice mismatch to suppress periodic energy barriers, but real interfaces invariably contain defects, edges and grain boundaries that restore high-friction states. Here we introduce a materials-agnostic strategy based on amorphous/crystalline heterointerfaces to achieve robust superlubricity under engineering-relevant conditions. Using diamond-like carbon (DLC) and crystalline $MoS_2$ as a model system, we show through experiments and atomistic simulations that their interface remains incommensurate at all orientations and exhibits vanishing energy barriers during friction. In contrast, twisted $MoS_2$ bilayers readily reorient into commensurate, high-friction states. We scale this effect by fabricating laser-patterned arrays of DLC/$MoS_2$ meta-contacts reinforced with $Ti_3C_2T_x$ MXene, forming hierarchical interfaces that sustain a friction coefficient of ~0.008 over 100000 cycles under combined extreme conditions: millimetre-scale contact size, 12.7 GPa contact pressure and RH 40% air. This unprecedented performance arises from four synergistic factors: intrinsic incommensurability at amorphous/crystalline interface, the rigidity of DLC support, MXene-based mechanical reinforcement and normalized load distribution by geometric patterning. These findings establish a general design paradigm that extends structural superlubricity from nanoscale model systems to practical technologies for sustainable engineering.

## 1. Introduction

Friction is one of the most pervasive energy-dissipating processes in modern technology, responsible for a substantial fraction of global energy losses,[1-3] and limiting the efficiency, precision and longevity of moving machinery.[4] Superlubricity,[5-7] a state of near-zero friction originating from the incommensurability effect,[8-10] has therefore attracted intense attention as a route to transform energy efficiency and mechanical reliability across scales.[11-12] Traditional solid superlubricity has been proved in homogeneous crystalline friction interfaces and heterogeneous crystalline friction interfaces. The former relies on twisting homogeneous interfaces (graphene/graphene,[13-17] graphite/graphite[18-22] and $MoS_2$/$MoS_2$[23]) to misfit angles, and the latter forms lattice mismatches between heterogeneous interfaces (graphene/$MoS_2$,[24-25] graphene/hexagonal boron nitride (*h*-BN),[26-30] graphite/*h*-BN,[31,32] graphite/gold,[33] graphene/gold,[34] $MoS_2$/Au,[35] and h-BN/carbon nanotubes[36]) with different lattice constants. Large-lattice-mismatch in heterogeneous interfaces avoids periodic energy barriers, but edge-pinning effect,[25,37,38] interfacial elasticity effects[39,40] and chemical contamination[41] induce high-friction states and cause the failure of superlubricity. Therefore, the necessities of achieving superlubricity require maintaining idealized conditions—an atomically flat, weak interaction, molecularly clean and rigid layer-by-layer structures. Although superlubricity has been realized at rigorously controlled defect-free[42] and lack-edge[38] crystalline interfaces under relatively low loads,[43,44] its practical applications are restricted to only nano- and microscale[31] as well as idealized environment.[45,46] Nearly three decades of research, engineering-grade (under combined extreme conditions: millimeter-scale contact size, contact pressure of GPa magnitude, practical operating environment *etc.*) superlubricity has not been realized,[5,6] primarily because the existence of edges/defects/grain boundaries in real crystalline becomes dominant with increasing contact size, especially in practical applications.[47-49] Extending ideal incommensurate superlubricity under conditions of macroscale contacts,[50-52] high-load,[53,54] and environmental effects[55,56]—particularly under such coupled conditions—requires overcoming these major challenges: inevitably disordered structures in macroscale contacts, interfacial deformations under high loads, and high humility sensitivity.[20,57] Firstly, macroscale contacts exhibit random and uncontrollable contact states.[50] Secondly, incommensurate contacts show strong twist-angle dependence in current homostructures. Although large-lattice-mismatch interfaces can reduce this dependence, high loads in macroscale contacts cause out-of-plane deformations, which disrupt lattice incommensurability.[27,58] Thirdly, in reactive environment,[56,58] inevitably grain boundaries and defects of macroscale crystals become energetic sites of tribochemical reactions.[59] Therefore, it is difficult to maintain sustained incommensurate contacts particularly under dynamic friction, and much less engineering-grade superlubricity remains unrealized for practical applications.

In this study, we find a special phenomenon that DLC/$MoS_2$ meta-contacts remain persistent incommensurate at all twist angles in view of amorphous DLC lacks of fixed lattice constant, anisotropy and grain boundaries. To further scale this effect to macroscale, a new concept of normalized meta-contact design is proposed. By patterning macroscale random contacts into controllable amorphous DLC/crystalline $MoS_2$ meta-contacts, regular arrays are constructed actively of hard DLC pillar coated with $MoS_2$ and reinforced MXene hoping to obtain engineering-grade superlubricity under combined extreme conditions: millimeter-scale contact size, contact pressure of GPa magnitude, practical operating environment etc., which will extend structural superlubricity from nanoscale into engineering applications.

## 2. Results and Discussion
### 2.1. Persistent incommensurate amorphous/crystalline contacts

To observe the contact configurations friction-induced between crystalline $MoS_2$/$MoS_2$ and amorphous DLC/crystalline $MoS_2$, the two contacts were prepared by rubbing a $MoS_2$ layer onto the surfaces of a $MoS_2$ layer and a DLC film, respectively. The configurations of these two systems were examined along the c-axis by high resolution transmission electron microscopy (HRTEM). We find that after friction, although $MoS_2$/$MoS_2$ exhibit incommensurate contacts upon twisting at 10° or 30°,[18] friction-induced commensurate contacts predominate across $MoS_2$/$MoS_2$, occurring at twist angles of 0° or 60° (I, II, and III in **Figure 1**a; Figure S1a, Supporting Information). In contrast, regardless of the relative twist angle of $MoS_2$ (110) and (100) lattice planes on DLC surface,[60] the DLC/$MoS_2$ contacts exhibit twist-angle-independent, referring to atomistic non-locking states (IV, V, and VI in Figure 1a; Figure S1b, Supporting Information).

The molecular dynamics (MD) simulations were constructed to correlate with experiments. We designed the initial sliding models of DLC/$MoS_2$ and $MoS_2$/$MoS_2$. The middle $MoS_2$ layers established incommensurate contacts with the outer $MoS_2$ layers by rotating 20°,[18] and behaviors of middle MoS$_2$ layers during dynamic friction process were observed (Figure S2, Supporting Information). With the proceeding of friction process, incommensurate contacts of MoS$_2$/MoS$_2$ gradually transitioned to commensurate states (Figure 1b, MoS$_2$/MoS$_2$; movie S1, Supporting Information), accompanied by a concurrent increase in friction that eventually stabilized. Relevant MD simulations reveal that the average friction force of MoS$_2$/MoS$_2$ is 1.08 nN in these commensurate states (Figure 1c). It is important to note that DLC/MoS$_2$ contacts remain mismatched at all twist angles and slide with vanishing energy barriers throughout the entire dynamic friction process (Figure 1b, DLC/MoS$_2$; Movie S2, Supporting Information). The corresponding MD simulations confirm DLC/MoS$_2$ contacts ultralow friction force of only 0.31 nN (Figure 1c), which is significantly lower than sliding between MoS$_2$/MoS$_2$ contacts.

Incommensurate contacts of MoS$_2$/MoS$_2$ are transient and rapidly transition toward commensurate contacts as friction progress, leading to high-friction states. On the contrary, DLC/MoS$_2$ maintains persistent incommensurate contact states. Moreover, DLC/MoS$_2$ contacts remain mismatched at all twist angles and slide with vanishing energy barriers, whereas MoS$_2$/MoS$_2$ bilayers reorient into high-friction commensurate states. As a result, the limitations of ultralow friction in crystalline materials depending on lattice constants and twist angles are overcome, and the influence of edge effects to crystalline materials is greatly weakened,[61] allowing DLC/MoS$_2$ contacts to avoid atomistic locking and achieve advantageous ultralow friction compared to MoS$_2$/MoS$_2$ contacts.

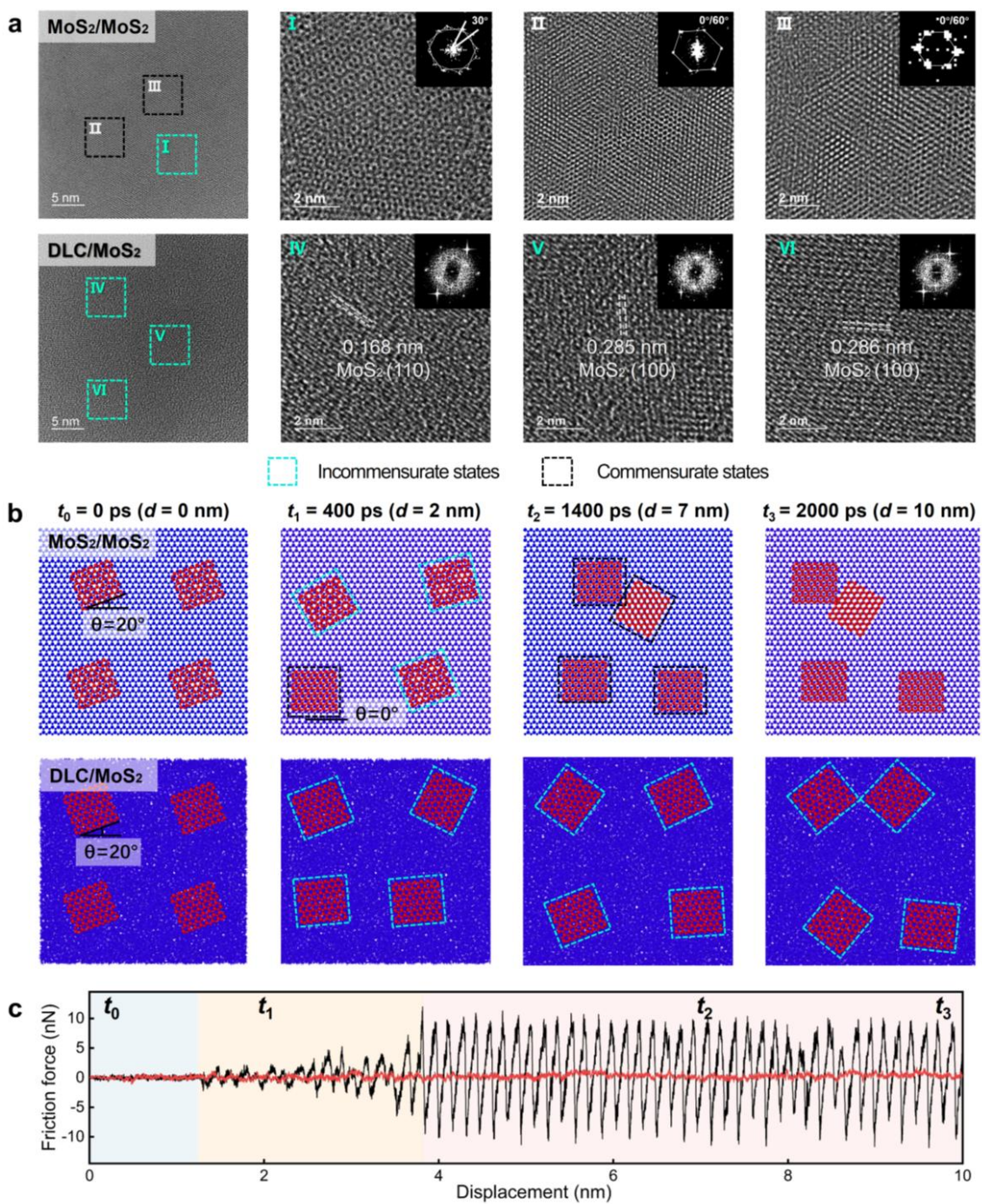

**Figure 1.** HRTEM images and MD simulations of crystalline $MoS_2/MoS_2$ and amorphous DLC/crystalline $MoS_2$ contacts. a) HRTEM images of $MoS_2/MoS_2$ and $DLC/MoS_2$ contacts. (I), (II) and (III) are the enlarged images of the marked positions in $MoS_2/MoS_2$ contacts. (IV), (V) and (VI) are the enlarged images of the marked positions in $DLC/MoS_2$ contacts. Blue and black squares mark the regions of incommensurate and commensurate states respectively, and the insets are fast Fourier transformation (FFT). b) MD simulations of $MoS_2/MoS_2$ and $DLC/MoS_2$ contacts at four different times ($t_0$=0 ps, $t_1$=400 ps, $t_2$=1400 ps and $t_3$=2000 ps). c) Friction force-displacement curves of $MoS_2/MoS_2$ and $DLC/MoS_2$ contacts at four different times.

## 2.2. Design of normalized DLC/MoS₂ meta-contacts

Extending the above persistent incommensurate $DLC/MoS_2$ contacts to macroscale presents a key challenge where the macroscale contacts are random and disordered (**Figure 2**a), causing the different contact area, shape, stress distribution, orientation, and chemical structure of each asperity exert different effects on friction. A normalized contact design is proposed to obtain the same controllable contact state, and macroscale contact is constructed by amorphous carbon/crystalline meta-contacts through three steps to avoid the adverse influence of random contacts on macroscale. Firstly, patterned meta-contacts were obtained by dividing macroscale contact surface *via* laser texture (Figure 2b; Figure S3a, Supporting Information). Secondly, amorphous DLC film was deposited on each meta-contact by magnetron sputtering (Figure 2b; Figure S3b, Supporting Information), and original structure of DLC is shown in Figures S4b and S8. Thirdly, $MoS_2$ and MXene were sprayed on the surface of amorphous DLC film by spraying method (Figure 2b; Figure S5, Supporting Information). The friction tests show that engineering-grade superlubricity (friction coefficient is about 0.009 in Figure S6, Supporting Information) is obtained with the constructed DLC/$MoS_2$/MXene composite on textured steel in 40% RH under average contact pressure of 12.7 GPa (Figure S7, Supporting Information) and millimeter-scale contact. The wear scar and wear track, at superlubriciting stage (30 min) show that crystalline $MoS_2$/MXene composite powder is transferred to the wear scar as shown in scanning electron microscope (SEM) morphology (I in Figure 2c; Figure S13, Supporting Information) and time of flight secondary ion mass spectrometry (TOF-SIMS in Figure S12a, Supporting Information). It indicates that the wear track consists of amorphous DLC film from meta-contact surfaces and crystalline composite powder filling the grooves between meta-contacts (II in Figure 2c; Figure S13, Supporting Information). Additionally, TOF-SIMS images demonstrated a regularly arranged array at the wear track, and this configuration extends in three dimensions, forming a well-defined topological structure (Figure S12b, Supporting Information). The $I_D$/$I_G$ intensity ratio of DLC unchanged after friction (Figures S5 and S13, Supporting Information), indicating that no structural transformations during friction. In other words, the DLC structure on meta-contacts surface remains identical to pristine film after friction. It demonstrates that macroscale friction primarily occurs at the interface of amorphous DLC/crystalline $MoS_2$. In this way, the DLC/$MoS_2$ meta-contacts is successfully scale up to macroscale by patterning macroscale contacts into regular arrays, ultimately achieving robust macroscale superlubricity in the DLC/$MoS_2$/MXene composite on textured steel.

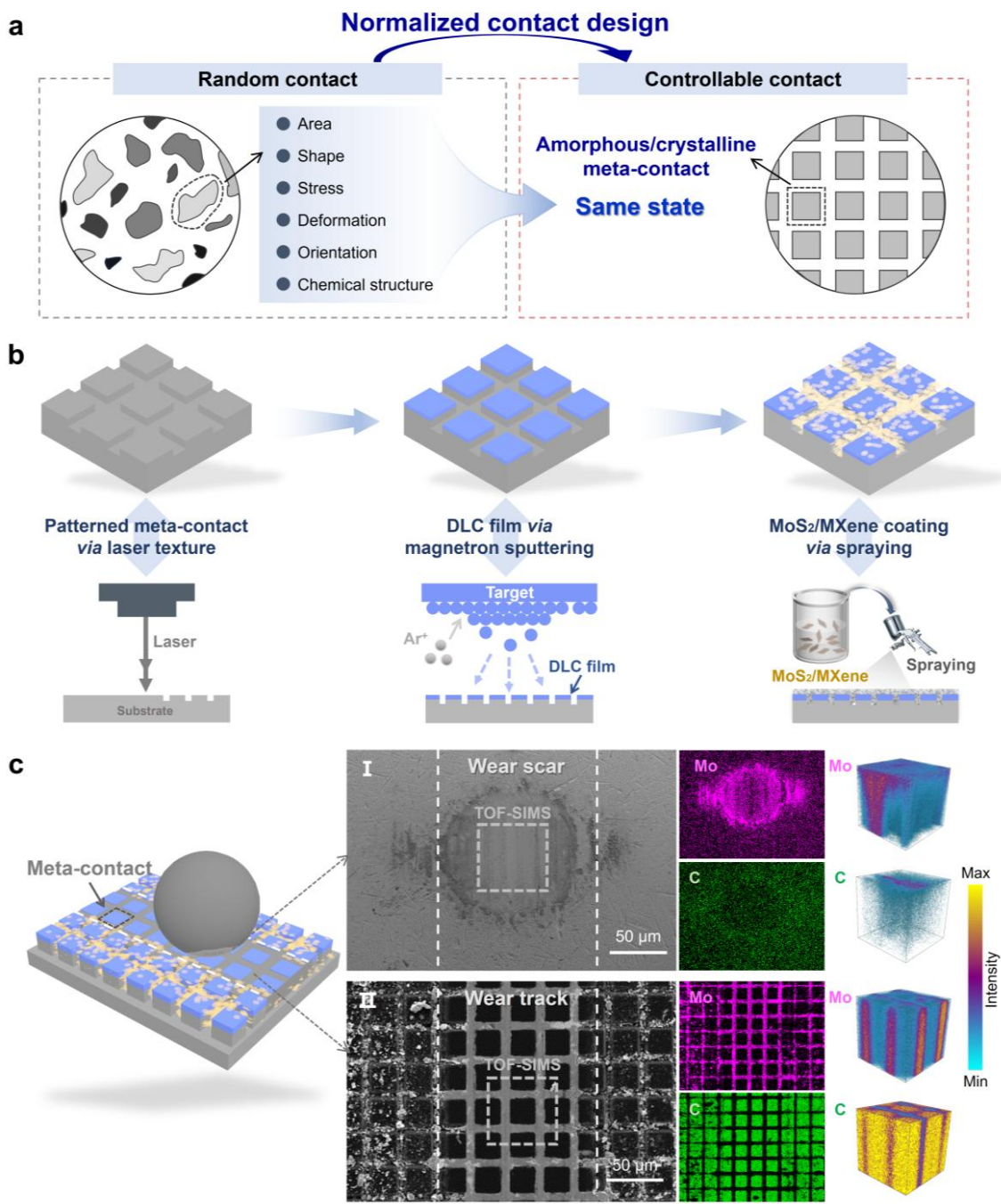

**Figure 2.** Design and construction of DLC/MoS$_2$ meta-contact interface. a) The schematic of normalized contact principle. b) Three steps for constructing amorphous DLC/crystalline MoS$_2$ meta-contact. c) The morphologies of friction interface contained meta-contacts. SEM morphology, EDS mappings and TOF-SIMS images of wear scar (I) and wear track (II). The intensity represents the ratio of mass-to-charge (m/z).

## 2.3. Suppressing deformation under extreme high contact pressure

To further investigate the roles of DLC and MXene in the DLC/MoS$_2$/MXene composite on textured steel, three comparative systems were designed (MoS$_2$ coating on textured steel, DLC/MoS$_2$ composite on textured steel, and DLC/MoS$_2$/MXene composite on textured steel) under extreme high load conditions. Their tribological properties and micromorphology of friction interfaces show that MoS$_2$ coating on textured steel exhibited a high friction coefficient (approximately 0.02) and a short wear life (around 12,500 laps) under average contact pressure of 12.7 GPa. Corresponding HRTEM images of the wear track show sever out-of-plane deformations, including broken and bent MoS$_2$ layered structures as well as even deformed steel substrate (**Figure 3**a; Figure S14, Supporting Information). The results indicate that both steel substrate and MoS$_2$ layered structures undergo severe damage under extreme high contact pressure, which accounts for high friction coefficient and short wear life of MoS$_2$ coating on textured steel.

The DLC/MoS$_2$ composite on textured steel achieves macroscale superlubricity under identical experimental conditions, exhibiting a friction coefficient about 0.009, albeit with a superlubricity lifespan of approximately 12,500 laps. Relevant HRTEM images of the wear track indicate that MoS$_2$ at the friction interface exhibits bent layered structures (Figure 3b), referring to subsequent increase in friction and the eventual failure of superlubricity. The results indicate that the random network structure of $sp^3$/$sp^2$ bonded carbon in the amorphous DLC of DLC/MoS$_2$ contacts mitigates high friction coefficient caused by steel substrate deformations and enables short-term superlubricity. However, MoS$_2$ layered structures undergo severe bending under high contact pressure in air. Notably, the DLC/MoS$_2$/MXene composite on textured steel system achieves robust engineering-grade superlubricity with a friction coefficient of approximately 0.008 and an extended wear life of more than 18000 laps (Figure 3c), under the same experimental conditions. Relevant HRTEM images reveal that MoS$_2$ at the friction interface exhibits well-defined and straight layered structures (VII and VIII in Figure 3c), and a distinctly stacked layered composite of MoS$_2$/MXene appears at the friction interface (IX in Figure 3c). The MoS$_2$/MXene composite forms an alternating multilayer architecture with ideally flat layered stacking. This is attributed to exceptional mechanical strength and hardness of MXene, which retain its structural integrity under extreme high contact pressure.[62] During friction, MXene provides reinforcement and anchoring for MoS$_2$, thereby preserving the layered structures of MoS$_2$ and shielding it from severe mechanical stress and reactive atmospheric species. Furthermore, MXene facilitates the maintenance of a stable interface between MoS$_2$ and DLC during the friction process, which contributes to the robust engineering-grade superlubricity observed in DLC/MoS$_2$/MXene composite on textured steel.

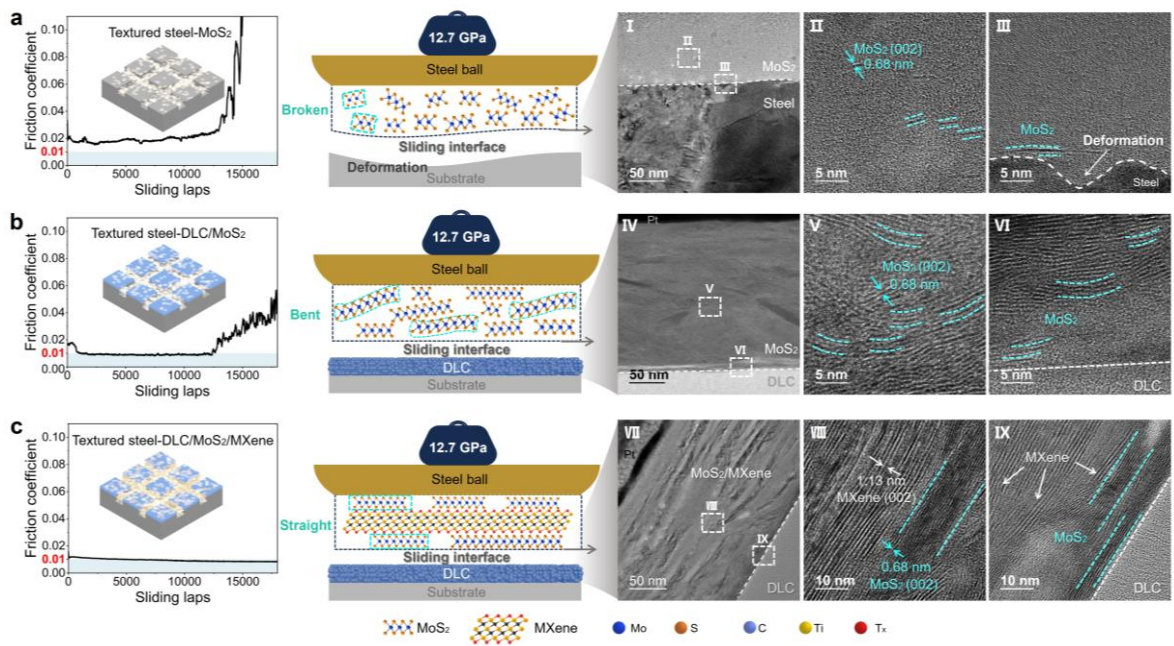

**Figure 3.** Friction curves and interface structures of different systems. a) The friction curve, friction interface diagram and HRTEM images of $MoS_2$ coating on textured steel (textured steel-$MoS_2$), (II) and (III) are the enlarged images of marked positions in (I). b) The friction curve, friction interface diagram and HRTEM images of DLC/$MoS_2$ composite on textured steel (textured steel-DLC/$MoS_2$), (V) and (VI) are the enlarged images of marked positions in (IV). c) The friction curve, friction interface diagram and HRTEM images of DLC/$MoS_2$/MXene composite on textured steel (textured steel-DLC/$MoS_2$/MXene), (VIII) and (IX) are the enlarged images of marked positions in (VII). The above friction tests were performed under 20% RH in air at average contact pressure of 12.7 GPa. The samples for HRTEM were selected after about 10,000 laps of friction (the stable friction stage).

## 2.4. The realization of engineering-grade superlubricity and relevant mechanisms

Scaling up DLC/$MoS_2$ meta-contacts to the macroscale contact level realizes engineering-grade superlubricity under coupled conditions of millimeter-scale contact, 40% RH in air, average contact pressure of 12.7 GPa, and a maximum linear velocity of 10 cm/s (namely frequency is 6.37 Hz). An engineering-grade superlubricity (friction coefficient of 0.008) and an ultralong wear life of above $1\times10^5$ cycles are exhibited in this system, meanwhile, maintaining robust macroscale superlubricity across a wide range of loads and 30% RH in $N_2$ atmosphere (**Figure 4**a; Figures S15 and S17, Supporting Information). Primarily, the DLC/$MoS_2$ contact configurations exhibit an inherent superlubricity advantage, not limited by periodic energy barriers. Subsequently, the topological arrangement of DLC/$MoS_2$ meta-contacts enables to the extension of superlubricity to macroscale. As a result, friction occurs between $MoS_2$ on the wear scar of steel ball and DLC on the meta-contacts, while the grooves between meta-contacts are filled with crystalline lubricating materials, which is continuously replenished at the friction interface to maintain engineering-grade superlubricity.

The design integrates from both constructing amorphous/crystalline contacts and normalized meta-contacts, enabling engineering-grade superlubricity in macroscale consisting of amorphous DLC/MoS$_2$ meta-contacts. This hierarchical interface is constructed in a stepwise design from the macroscale down to the microscale (Figure 4b). Firstly, normalized meta-contact arrays are constructed to form the macroscale contact, thereby obtaining the same contact states (I in Figure 4b), and avoiding the influence of adverse factors such as shape, orientation and deformation of contact points. Secondly, rigid DLC (II in Figure 4b) with exceptional strength and hardness ensures that the substrate maintains its structural integrity even during friction under extreme high contact pressure.[42] Thirdly, MXene as a reinforcement and anchoring agent is incorporated to ensure that MoS$_2$ maintains its ideal layered structures despite of extreme high contact pressure and atmospheric environment, which also imparts humidity insensitivity to MoS$_2$ and thus retaining the superlubricity interface of DLC/MoS$_2$ (III in Figure 4b). Finally, utilizing the atomic lattice mismatch design of DLC/MoS$_2$ contacts, the disordered amorphous DLC structure formed permanent incommensurate contact states with the ordered MoS$_2$ structure, eliminating the atomistic locking caused by lattice match (IV in Figure 4b). Such a comprehensive design covering the abovementioned perspectives contributes to achieve the robust engineering-grade superlubricity in atmosphere under extreme high contact pressure.

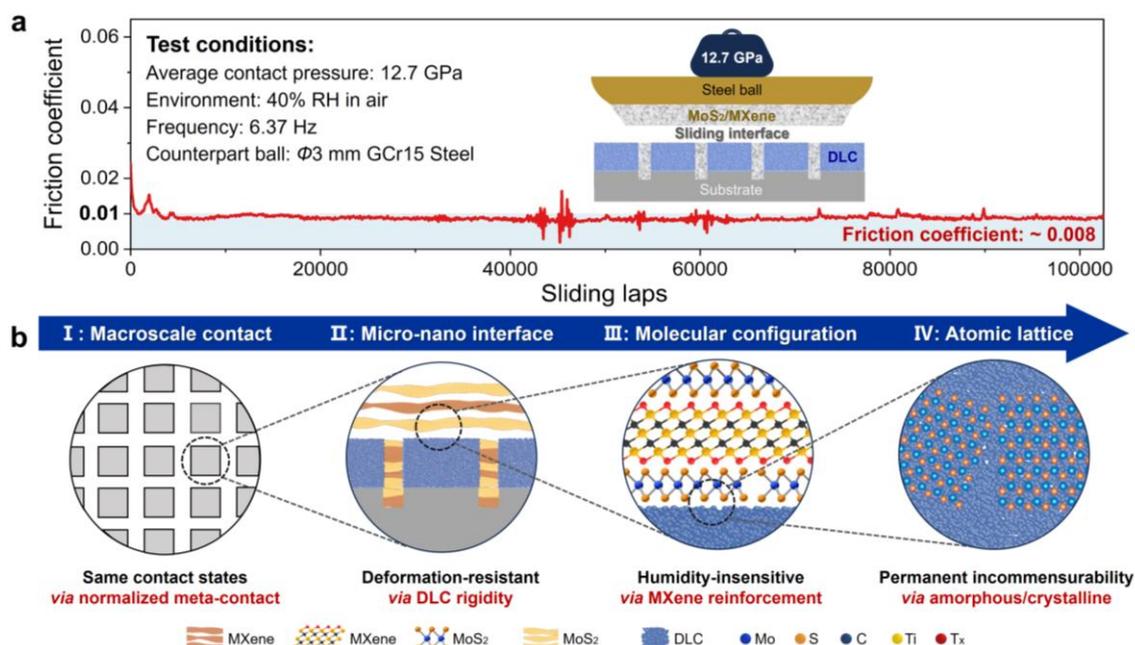

**Figure 4.** Superlubricaty and relevant mechanisms. a) Superlubricating curve of DLC/MoS$_2$/MXene composite on textured steel. b) Relevant hierarchical superlubricating mechanisms of DLC/MoS$_2$/MXene composite on textured steel from macroscale contact to atomic lattice scale. The schematic of (I) obtaining same contact states *via* normalized meta-contact, (II) deformation-resistant *via* rigid amorphous DLC, (III) humidity-insensitive *via* MXene reinforcement, and (IV) permanent incommensurability *via* amorphous/crystalline atomic lattice.

## 3. Discussion

In this study, combining experiments and simulations, a special phenomenon that DLC/MoS$_2$ meta-contacts maintains persistent incommensurate and energy barriers vanishes at all twist angles during the dynamic macro-friction is found, because DLC possessing inherent amorphous characteristics (lacking a fixed lattice constant, anisotropy and grain boundaries). Furthermore, DLC exhibits exceptionally high rigidity and withstands engineering-grade loads without out-of-plane deformation. The design concept of normalized meta-contact, which ingeniously resolves the major challenge of macroscale random contact, is significant for achieving engineering-grade superlubricity. Accordingly, numerous regularized amorphous carbon/crystalline meta-contacts were constructed and assembled into regular arrays to form macroscale contact. The robust superlubricity coupling with size, force, and environment under engineering-grade conditions was experimentally achieved (contact area of millimeter scale; average contact pressure of 12.7 GPa; 40% RH in air; friction coefficient: 0.008; lifetime > 1 × 10$^5$ laps), being the representative of real engineering-grade superlubricity.

This study breaks the long-standing reliance on energy dissipation caused by lattice matching in superlubricity, opening up new pathways in amorphous tribology. It offers a universal strategy to address the fundamental challenge of random macroscale contacts, thereby establishing a solid foundation for the engineering application of superlubricity technologies in fields such as aerospace, high-precision manufacturing, and transportation.

## 4. Experimental Section

*Materials*: Crystalline MoS$_2$ powder (size lower than 10 μm) was purchased from Benxi Xinliyuan Metal Materials Company. (Liaoning, China). Crystalline MXene (Ti$_3$C$_2$T$_x$) powder (size 2~10 μm) was purchased from XFNANO Technology Company. (Jiangsu, China). Analytical grade C$_2$H$_5$OH was commercially obtained. The high speed steel substrate (M2, Figure S4, Supporting Information) was purchased from Suzhou Feiyue Precision Machinery Company Limited (Jiangsu, China). Commercial Polyamide imide (PAI) was purchased from Beijing Zhongfu Technology Company Limited.

*Sample preparation*: The substrates included initial and textured steel disks with an ultrashort pulse laser surface etching system (Laser fine processing equipment, AS-5680, Jiangyin Deli Laser Equipment Company Limited, China; the energy and power of the laser were maintained at 8 kW and 10%) to obtain the laser textured steel disks. MoS$_2$ powder and MoS$_2$/MXene binary composite powder (mass ratio 1:1) were dispersed in ethanol at a concentration of 1 g·L$^{-1}$. The as-obtained dispersions were sonicated for 2 h. During the spraying process with high-purity nitrogen (99.9%) as the carrier gas, the pressure was maintained at about 0.2 MPa. Upon completion of spraying, the as-sprayed samples were kept under dry environment with less than 10% RH for 24 h to remove residual solvent. We also prepared MoS$_2$/MXene/PAI bonded solid lubricating coating, 0.5 g of MoS$_2$ and MXene (mass ratio 1:1) as well as 0.3 g of PAI resin with a solid content of 33% were added to 40 mL of dimethylformamide solvent. The resultant suspension was ultrasonically treated for 2 h and sprayed onto the surface of the DLC film on textured steel. The sprayed samples were then cured. The DLC film was deposited on the surface of the initial and textured steel disks by magnetron sputtering. The details of the DLC film preparation process are available in previous work.[63]

*Frictional Tests*: All the friction tests were conducted on a ball-on-disk tribometer (CSM Instrument, Switzerland) whose chamber gas environment can be readily controlled. Different ratios of high-purity air and water vapor were vented into the chamber of tribometer to regulate the relative humidity (RH) of test environment. The coated steel substrate disks were driven to slide against the steel counterpart ball ($\Phi$3 mm, AISI52100, $R$a ≈ 20 nm, Shanghai Steel Ball Factory Company Limited; Shanghai, China). The tests were conducted under linear reciprocating mode at a full amplitude of 5 mm, frequency of 6.37 Hz (max linear speed: 10 cm·s$^{-1}$), a applied load of 20 N (average contact pressure: 12.7 GPa; Figure S7, Supporting Information), and a room temperature of 23~25 °C. Before the start of the friction tests, the samples were dried at 40°C in vacuum (<0.1 Pa) for 1.5 h to evaporate ethanol and H$_2$O. The RH of the tribometer chamber was kept at 20%. To obtain the superlubricity properties under combined extreme conditions: millimetre-scale contact size, average contact pressure of 12.7 GPa and RH 40% air, friction test of prepared MoS$_2$/MXene/PAI bonded solid lubricating coating was carried out. All friction tests were repeated three times at least.

*Characterizations*: The Raman spectra (HORIBA JobinYvon S.A.S, LabRAM HR Evolution, France) were recorded at a wave length of 532 nm (2.3 eV). A JEOL-2010 HRTEM (Joel Corporation, Japan) was performed to analyze the microstructure of the original asperity surfaces, the wear scars and the wear tracks. Further, the wear tracks of the MoS$_2$ coating on textured steel, DLC/MoS$_2$ composite on textured steel and DLC/MoS$_2$/MXene composite on textured steel were cut with a focused ion beam (FIB; FEI, Helios 600, USA) to obtain the cross-sections.[39] Before the process of Ga$^+$ ion cutting, Pt layer was deposited on the surface of target sample to protect it from possible damage. The microstructure of wear scars and wear tracks was analyzed by HRTEM (FEI, TECNAI G2

S-TWINF20, USA). The 3D profiles of texture steel and DLC film on textured steel were described with a MicroXAM-800 3D surface profiler (KLA-Tencor, USA). The surface morphologies of the prepared coatings were observed with field-emission scanning electron microscope (FESEM, JSM-6701, Japan), and the TOF-SIMS was conducted with an IONTOF-TOF-SIMS M6 facility (German) (sputtering area: 300 μm × 300 μm; sputtering depth: 300 nm).

*MD Simulations*: MD simulations were performed to further unveil the suppression of friction-induced atomic interlocking effects by amorphous carbon. The MD simulations in this work were performed with the LAMMPS simulation package.[64] The interlayer interaction was described by the anisotropic interlayer potential (ILP) for H-diamond and $MoS_2$, and the intralayer interactions of DLC and $MoS_2$ were described by the second-generation REBO potential[65] and Stillinger–Weber (SW)[66] potential, respectively. The specific model setup is provided in Figure S2. The DLC model was generated through a melt-quenching process following the previous work.[61] Afterward, the unsaturated C atoms were passivated by hydrogen atoms. The DLC/$MoS_2$ ($MoS_2$/$MoS_2$) friction pair consists of three parts, with the bottom periodic DLC ($MoS_2$) layer fully fixed, the top periodic DLC ($MoS_2$) set as a rigid body, and the middle $MoS_2$ flakes as a free layer. At the beginning of the simulation, the middle $MoS_2$ flakes were rotated by 20° around the X axis to establish an incommensurate contact (Figure S2, Supporting Information). The Langevin thermostat[67] was applied to the $MoS_2$ flakes to maintain the system temperature at 300 K. External forces and velocities were applied to each atom in the rigid layer, where a applied load of 10 nN was imposed, and the rigid layer was driven to slide at a velocity of 5 m·s$^{-1}$. All visualization was performed using the OVITO software.[68]

**Supporting Information**

Supporting Information is available from the Wiley Online Library or from the author.

**Acknowledgements**

This work was supported by the Strategic Priority Research Program of the Chinese Academy of Sciences (No. XDB0470202), the National Natural Science Foundation of China (Nos. 52275222, 52405232, 12472099 and U2441207), the Fundamental Research Funds for the Central Universities (Nos. 2042025kf0050, 2042025kf0013 and 600460100), and Longyuan Youth Talent Project. Computations were conducted at the Supercomputing Center of Wuhan University and the National Supercomputer TianHe-1(A) Center in Tianjin.

**Conflict of Interest**

The authors declare no conflict of interest.

**Data Availability Statement**

The data that support the findings of this study are available in the supporting information of this article.

# Supporting Information

## Persistent Incommensurate Amorphous/Crystalline Meta-Interfaces Enable Engineering-Grade Superlubricity

*Wan Wang, Zijun Ding, Panpan Li,\* Wanying Ying, Hongxuan Li, Xiaohong Liu, Huidi Zhou, Jianmin Chen, Wengen Ouyang,\* and Li Ji\**

W. Wang, P. Li, H. Li, X. Liu, H. Zhou, J. Chen, L. Ji
State Key Laboratory of Solid Lubrication, Lanzhou Institute of Chemical Physics, Chinese Academy of Sciences, Lanzhou 730000, P. R. China
E-mail: lipanpan@licp.cas.cn; jili@licp.cas.cn.

Z. Ding, W. Ying, W. Ouyang
Department of Engineering Mechanics, School of Civil Engineering, Wuhan University, Wuhan, Hubei, 430072, P. R. China
E-mail: w.g.ouyang@whu.edu.cn

W. Ouyang
State Key Laboratory of Water Resources Engineering and Management, Wuhan University, Wuhan, Hubei, 430072, P. R. China

W. Wang, H. Li, X. Liu, L. Ji
Center of Materials Science and Optoelectronics Engineering, University of Chinese Academy of Sciences, Beijing 100049, P. R. China
W. Wang and Z. Ding contributed equally to this work

**This PDF file includes:**
   Supplementary Text
   Figs. S1 to S17

   **This Supporting Information includes the following sections:**
   1. Interfacial structure evolution of $MoS_2/MoS_2$ and $DLC/MoS_2$ during sliding
   2. MD simulation setup for $MoS_2/MoS_2$ and $DLC/MoS_2$ heterostructures
   3. Morphologies and structure of the original materials
   4. Superlubricity of $DLC/MoS_2/MXene$ composite on textured steel
   5. Finite element method (EFM) of interface contact pressure
   6. Structure and composition of the original materials
   7. The friction interface of $DLC/MoS_2$
   8. Severe out-of-plane deformation of $MoS_2$ coating on textured steel
   9. Tribological properties of $DLC/MoS_2/MXene$ composite on textured steel and DLC film

**Other Supporting Information for this manuscript include the following:**

   Movie S1: Transition of $MoS_2/MoS_2$ contact configurations during dynamic sliding.
   Movie S2: Transition of $DLC/MoS_2$ contact configurations during dynamic sliding.

**1: Interfacial structure evolution of MoS₂/MoS₂ and DLC/MoS₂ during sliding**

Figure S1 shows the HRTEM images of crystalline MoS₂/MoS₂ and amorphous DLC/crystalline MoS₂ contact configurations. DLC film was deposited on the surface of KBr salt, and a layer of crystalline MoS₂ powder was wiped on the surface of DLC film. The results show that at the interface of MoS₂/MoS₂, torsion angle of MoS₂/MoS₂ interface is 0° or 60° at lots of locations, and the interface is in lattice matching commensurate. While all the contacts between DLC/MoS₂ are incommensurate regardless of the torsion angle of MoS₂ (100) or (110) crystal plane on the surface of DLC film.

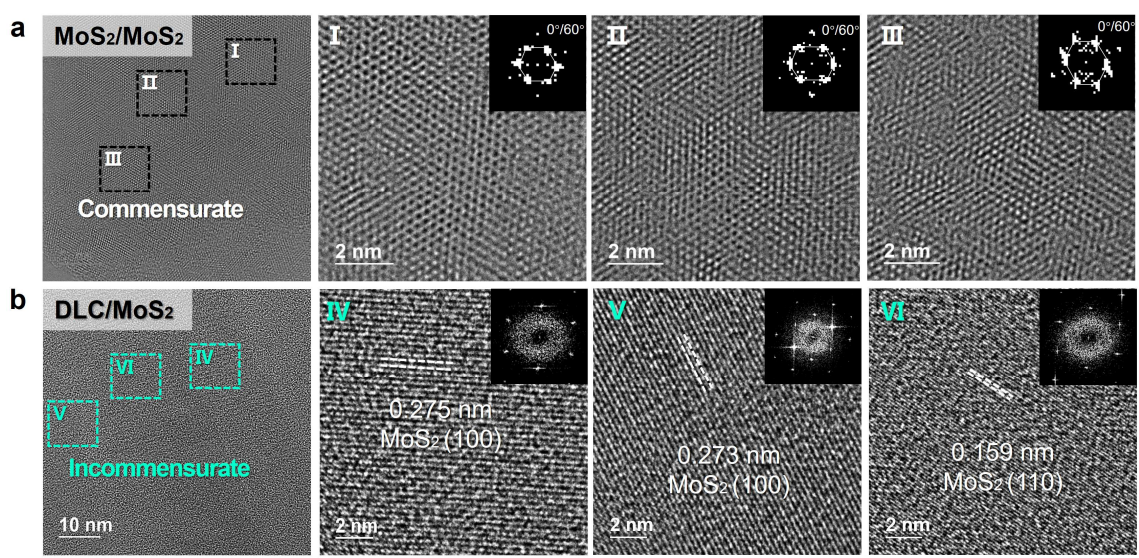

**Figure S1.** a) HRTEM images of MoS₂/MoS₂ contacts, (I), (II) and (III) are the enlarged images and FFT images of the marked positions in (a). b) HRTEM images of DLC/MoS₂ contacts, (IV), (V) and (VI) are the enlarged images and FFT images of the marked positions in (b).

## 2: MD simulation setup for MoS$_2$/MoS$_2$ and DLC/MoS$_2$ heterostructures

As shown in Figure S2, the DLC/MoS$_2$ friction pair consists of three layers, the top and bottom layers being periodic DLC structures and the middle composed of four aperiodic MoS$_2$ flakes. The periodic DLC structure has dimensions of 10.3 × 10.3 × 1.2 nm$^3$ and consists of 21913 C atoms and 3881 H atoms. Each aperiodic MoS$_2$ flake has dimensions of 2 × 2 × 0.3 nm$^3$ and is composed of 60 Mo atoms and 120 S atoms.

The MoS$_2$/MoS$_2$ friction pair also consists of three layers, where the top and bottom layers are periodic MoS$_2$ structures and the middle layer is composed of the same four aperiodic MoS$_2$ flakes as in the DLC/MoS$_2$ friction pair. The periodic MoS$_2$ structure has dimensions of 10.9 × 10.8 × 0.3 nm$^3$ and consists of 1400 Mo atoms and 2800 S atoms. The top periodic DLC (or MoS$_2$) layer was set as a rigid body and driven to move at 5 m·s$^{-1}$. The bottom periodic DLC (or MoS$_2$) layer was fully fixed.

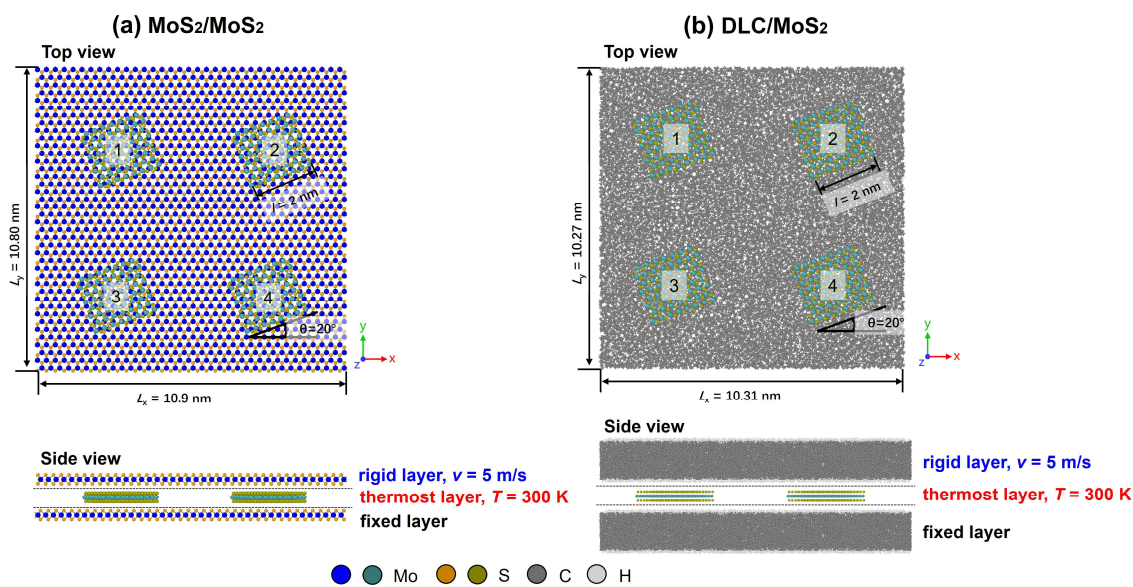

**Figure S2.** The model of MD simulations. a) MoS$_2$ versus MoS$_2$. b) DLC versus MoS$_2$.

## 3: Morphologies and structure of the original materials

The constructed meta-contact with a length of 28 μm, a width of 28 μm, and a height of 2.0 μm on the substrate surface by laser texture, and the spacing of two meta-contacts was 10 μm (Figure S3a). The DLC film was deposited on the textured steel substrates, obtaining DLC film on textured steel, and the height of the meta-contacts is 1.5 μm, the length and width of each meta-contact, and the spacing between the meta-contacts remain unchanged (Figure S3b).

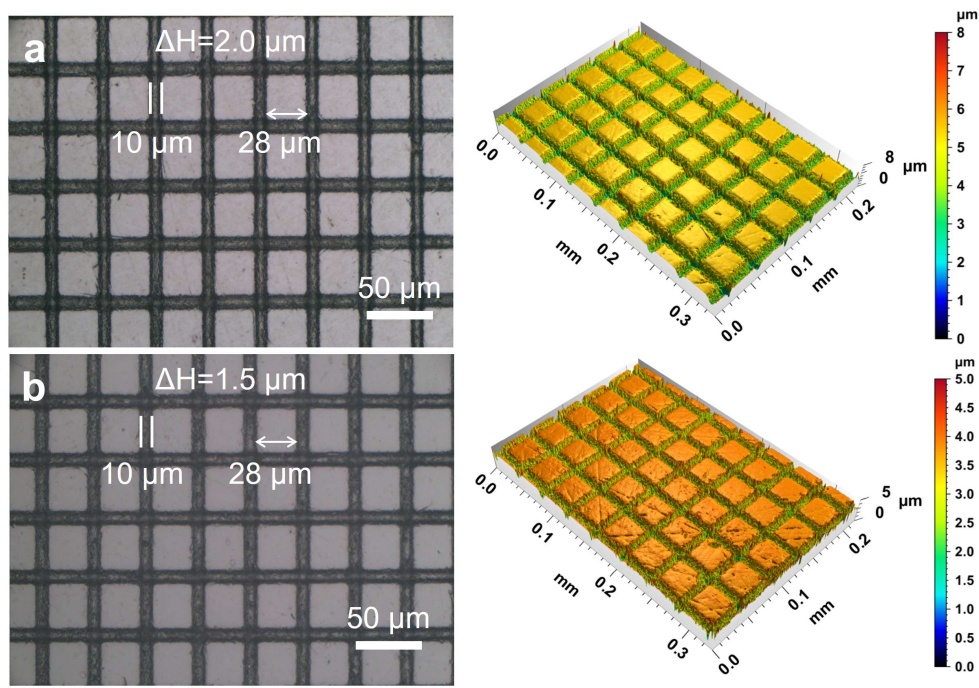

**Figure S3.** Morphologies and three-dimensional (3D) profiles of textured steel and DLC film on textured steel. a) textured steel, b) DLC film on textured steel.

The original steel substrate used in the experiment exhibits a roughness of 17 nm (Figure S4a), and the roughness is about 11 nm (Figure S4b) after the DLC film is deposited on original steel substrate.

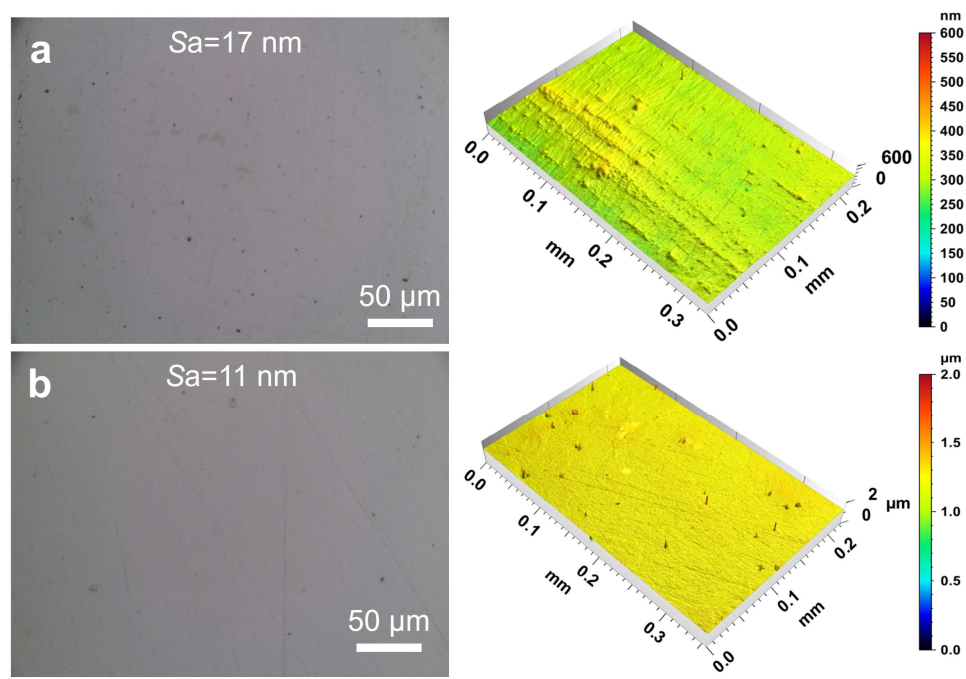

**Figure S4.** Morphologies and 3D profiles of original steel and DLC film. a) smooth steel, b) DLC film on smooth steel.

SEM morphologies and EDS mappings of the DLC/MoS$_2$/MXene composite on the textured steel show that MoS$_2$/MXene sheets are deposited uniformly on the surface of this sample (Figure S5a). Raman spectra show that MoS$_2$/MXene composite powder is deposited in the grooves and on the meta-contacts surface (Figure S5a, I and II), and the $I_D/I_G$ of original DLC fim is 0.78. The TOF-SIMS images also confirm the DLC film coverage the meta-contacts surface and the MoS$_2$/MXene composite distributes across both the meta-contact surfaces and grooves. (Figure. S5b).

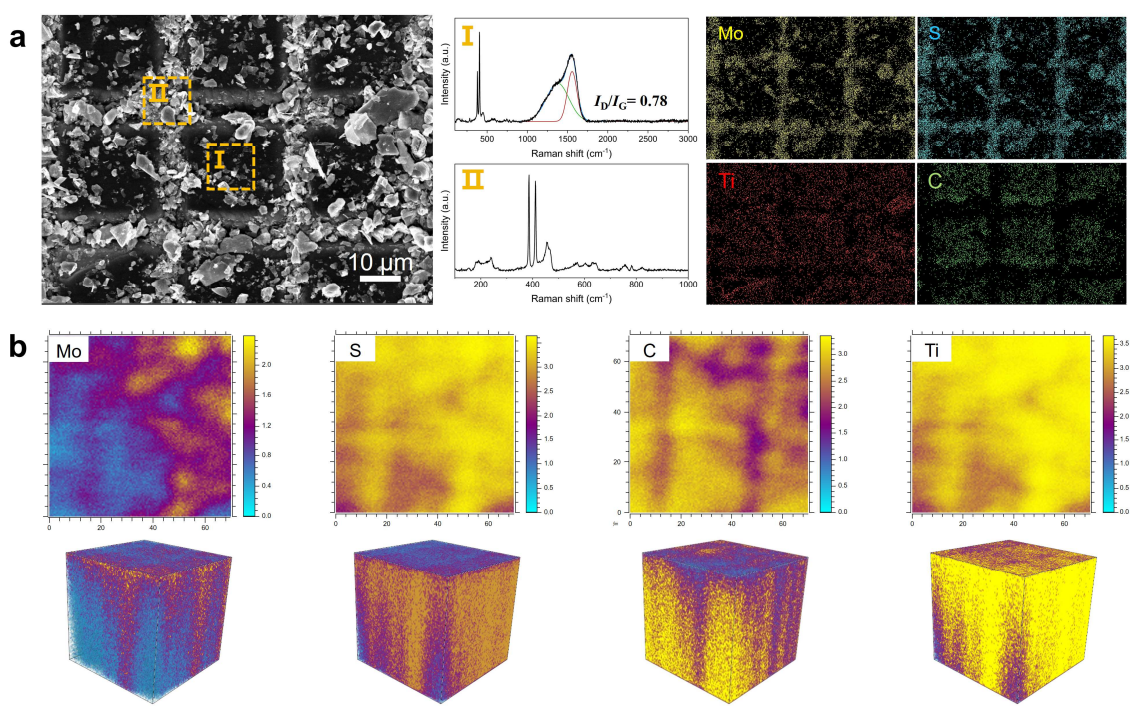

**Figure S5.** Surface morphologies, elemental distribution of DLC/MoS$_2$/MXene composite on textured steel. a) SEM morphologies and EDS mappings of DLC/MoS$_2$/MXene on textured steel, (I) Raman spectra of grooves, (II) Raman spectra of meta-contact surfaces. b) TOF-SIMS images of DLC/MoS$_2$/MXene on textured steel (Scanning depth: 300 nm, scanning area: 100 μm × 100 μm).

**4: Superlubricity of the DLC/MoS₂/MXene composite on textured steel**

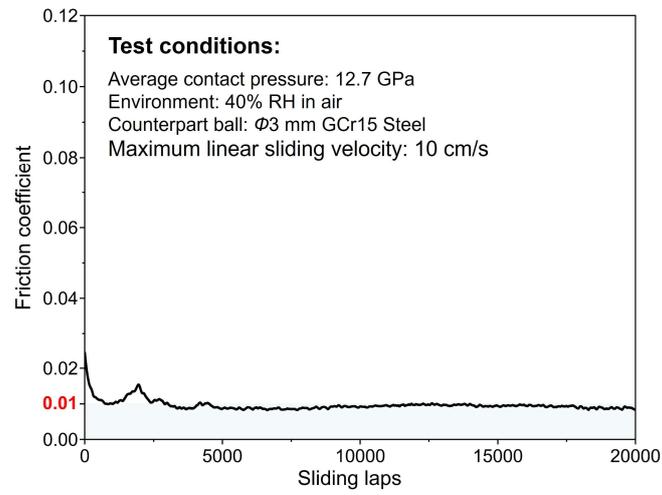

**Figure S6.** Friction curve of the DLC/MoS$_2$/MXene composite on textured steel.

**5: FEM of interface contact pressure**

The average contact pressure at the interface was calculated using the FEM. Based on experimental conditions, a corresponding computational model was established in the finite element software ABAQUS/Explicit. The model mainly consists of a hemispherical indenter and a DLC film (Figure S7a). The diameter of the hemisphere is consistent with the experiment, set at 3 mm. To improve computational efficiency, only the key contact region of the film was retained, with each block measuring 28 × 28 × 3 μm. Subsequently, material properties were assigned to each component according to experimental calibration, with specific parameters listed in Table S1.

To ensure computational stability, a quasi-static explicit algorithm was employed for solving the model. The contact interface between the bottom of the hemisphere and the DLC film was defined as surface-to-surface contact, with normal behavior set as "hard contact". The total simulation time was 1 s: a vertical concentrated load of 20 N was applied to the hemisphere in the first 0.5 s, and maintained for the remaining 0.5 s. Meanwhile, all degrees of freedom of the hemisphere except the loading direction were constrained, while the bottom of the film was fully fixed with symmetric constraints. A schematic of the model is shown in Figure S7a.

The simulation results show that the distribution of normal contact stress at the interface between the film and the hemisphere is illustrated in Figure S7b and S7c. The shape and size of the contact area agree well with experimental observations. During post-processing, CONRMF and CNAREA were selected to extract the normal contact force ($F_N$) and contact area ($A$), respectively. The average contact stress was then calculated using the formula $\sigma_N = F_N / A$.

To ensure model convergence while balancing computational efficiency and accuracy, the model uniformly used reduced-integration hexahedral elements (C3D8R), with local mesh refinement applied in the contact region between the hemisphere and the film. A mesh convergence analysis was performed for the DLC contact interface by setting mesh sizes of 0.004, 0.0035, 0.0032, 0.003, and 0.0028 mm sequentially. The contact stress over time in the same region was analyzed, and the results showed that the stress curves under different mesh sizes were nearly identical (Figure S7d), indicating that the results are mesh-independent. A similar analysis was conducted for the hemispherical contact surface, which also demonstrated mesh independence, thereby verifying the reliability of the simulation results.

Finally, a mesh size of 0.0028 mm was selected for the simulation. The entire contact region on the DLC interface was analyzed, yielding an average contact pressure of 12.7 GPa under the experimental conditions.

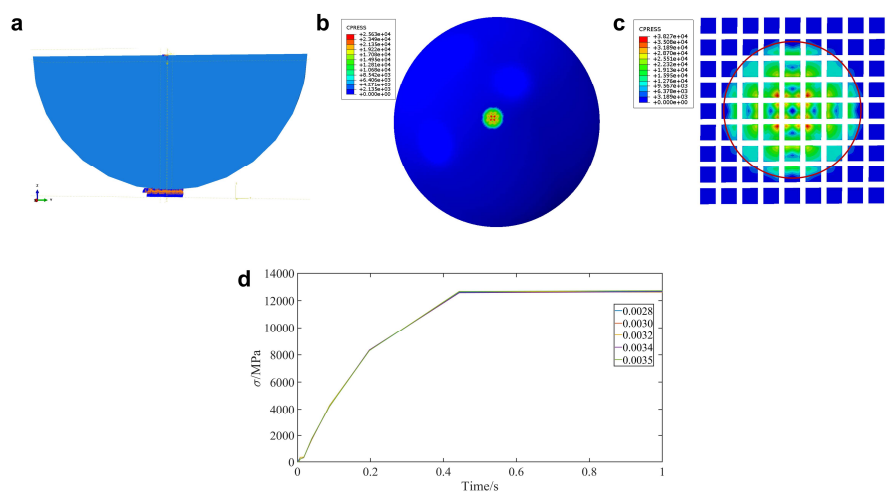

**Figure S7.** FEM of the average contact pressure at the interface. a) Schematic of the model. b) Contact stress distribution on the spherical surface. c) Contact pressure distribution on the DLC surface. d) Mesh convergence analysis.

**Table S1.** Materials parameters for the constructed model.

| Components of the model | Matrix | Modulus of elasticity (GPa) | Density (g/cm$^3$) | Poisson's ratio |
|---|---|---|---|---|
| **Ball** | GCr15 Steel | 208 | 7.81 | 0.3 |
| **Film** | DLC | 101.389 | 2.0 | 0.3 |

**6: Structure and composition of the original materials**

The hardness and elastic modulus of the DLC films are 11.25 GPa and 115.65 GPa,[1] respectively. The bonding strength between the DLC film and the high speed M2 steel substrate is 47.85 N. Figure S8 shows the scratch test morphology of the DLC. The critical load at which the film fails due to fracture is used as a criterion for adhesion. The position marked by the blue line is the initial peeling point (Lc1) of the film, and the peeling position corresponds to its first critical load.

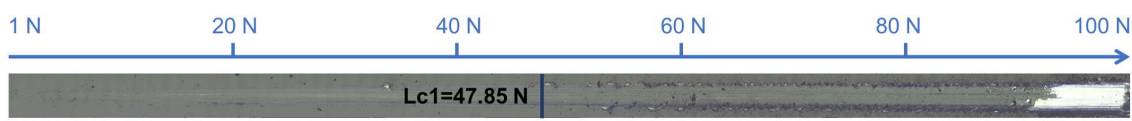

**Figure S8.** Scratch test morphology of the DLC film.

As shown in Figure S9, MXene and MoS$_2$ both are lamellar structures with the wrinkles at the edges. The MXene sheet is 9-10 layers, and the MoS$_2$ sheet is 6-7 layers.

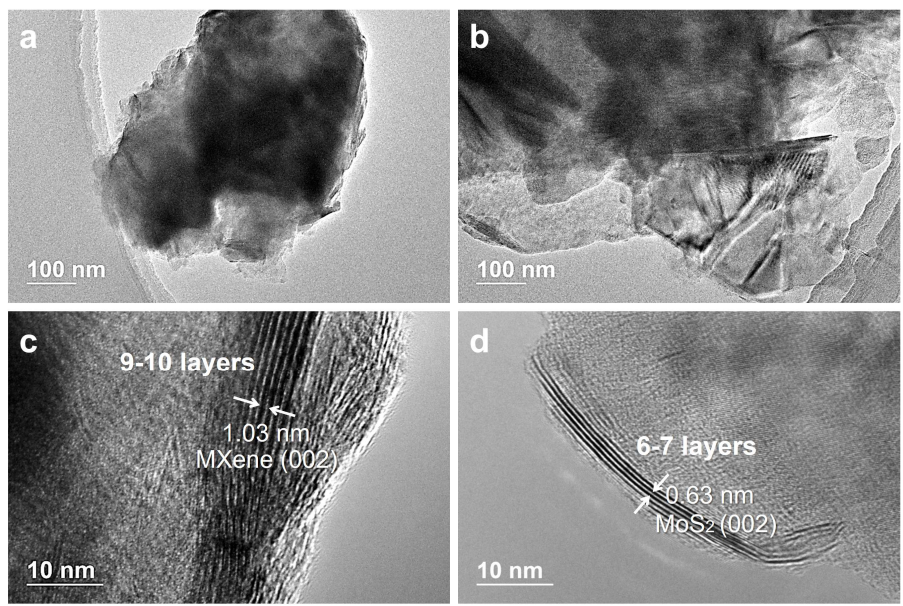

**Figure S9.** Morphologies of the original MXene and MoS$_2$ sheets. TEM morphologies of a) MXene, b) MoS$_2$. HRTEM images of c) MXene, d) MoS$_2$.

As shown in Figure S10a, the XRD pattern of MoS$_2$ shows that there is an evident diffraction peak at ~14° corresponding to the (002) crystal base plane of MoS$_2$. The XRD patterns of MXene shown in Figure S10b, there are diffraction peaks at ~9°, ~18° and ~27°, which are attributed to the (002), (004) and (006) crystalline planes of the MXene. Figure S10c is the Raman spectrum of MoS$_2$, the high interlayers S-Mo-S vibration of A$_{1g}$ peak is at about ~408 cm$^{-1}$, and the in-plane S-Mo-S vibration of E$^1_{2g}$ peak is at about ~384 cm$^{-1}$.[2] Figure S10d is the Raman spectrum of MXene, E$_g$ vibration of Ti, C, and T$_x$ groups is at about ~125 cm$^{-1}$, which is considered to the key characteristic peak for the successful synthesis of MXene. A$_{1g}$ (Ti, C and T$_x$) vibration is at about ~210 cm$^{-1}$ and another resonant Raman band at ~710 cm$^{-1}$, which corresponds to A$_{1g}$ (C) vibration.[3]

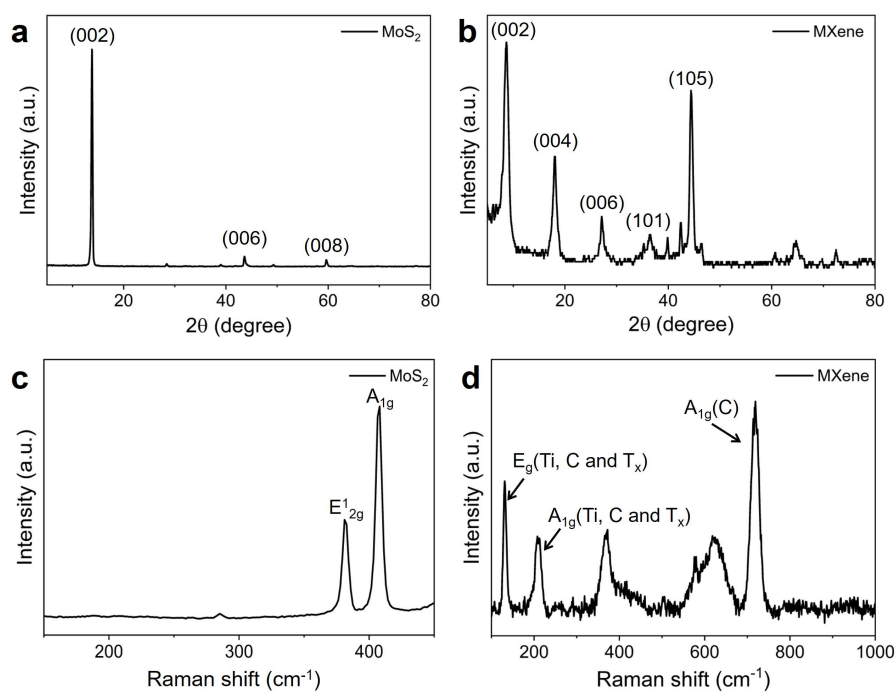

**Figure S10.** Structure and composition of the original MXene and MoS$_2$ sheets. XRD patterns of a) MoS$_2$, b) MXene. Raman spectra of c) MoS$_2$, d) MXene.

SEM morphologies of prepared coatings are shown in Figure S11. It can be seen the both MoS$_2$ and MXene sheets well spread at the substrate after spraying, and the lateral size of the MoS$_2$ sheets is about 2~5 μm and that of MXene sheets is about 2~10 μm Figure S11a and Figure S11b. When these sheets were composited to spray on the substrate, similarly the MoS$_2$ and MXene sheets stacked together and tiled cover the substrate (Figure S11c).

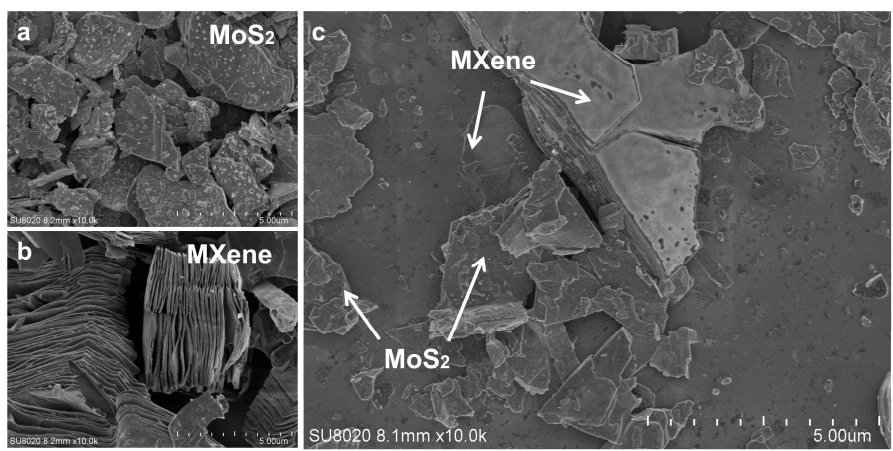

**Figure S11.** SEM morphologies of prepared coatings. a) MoS$_2$. b) MXene. c) MoS$_2$/MXene composite.

## 7: The friction interface of DLC/MoS$_2$

Figure S12 is TOF-SIM, and it shows the distribution of C, Ti, S and Mo elements in the two-dimensional (2D) direction and three-dimensional (3D) direction. The wear scar is mainly composed of Ti, S and Mo elements (Figure S12a), which indicates that the transfer film is mainly composed of MoS$_2$/MXene composite powder. That the wear track presents a regular topological structure (Figure S12b) in the 3D direction. It demonstrates that friction occurred at the interface of DLC and MoS$_2$/MXene.

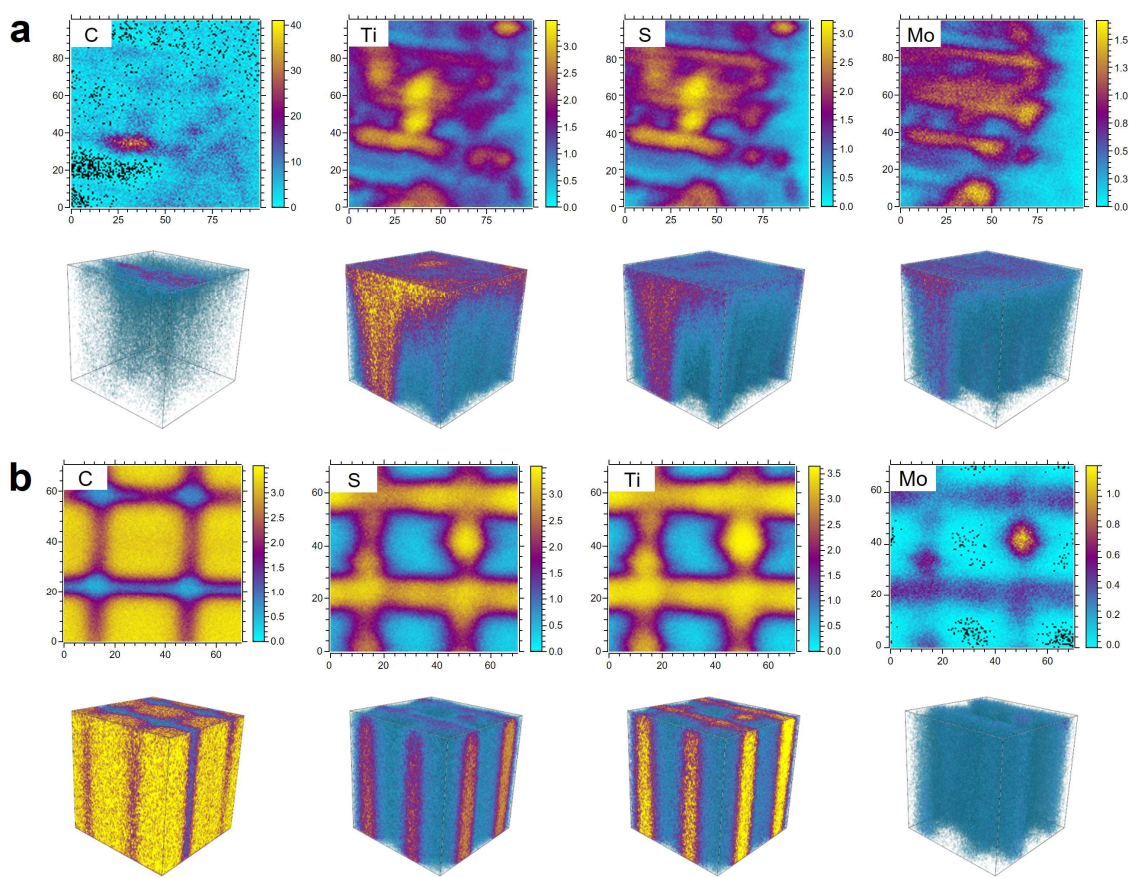

**Figure S12.** TOF-SIMS images of the friction interface. a) TOF-SIMS images of C, S, Ti and Mo element distribution in the 2D and 3D direction at the wear scar. b) TOF-SIMS images of C, S, Ti and Mo element distribution in the 2D and 3D direction at the wear track.

SEM images, EDS mapping and Raman spectra at the wear scar and wear track (Figure S13a, cI) show that the transfer film is composed of MoS$_2$ and MXene. There are two typical locations on wear track, and the meta-contact surfaces are covered by DLC film (Figure S13bIII, cIII), while the grooves between meta-contacts are filled with MoS$_2$ and MXene (Figure S13bII and cII). The $I_D/I_G$ of DLC film on the meta-contact surface before (0.78; Figure S5aI) and after (0.79; Figure S13CIII) friction is nearly unchanged, which indicates the structural integrity of the DLC film is maintained during the friction process, as evidenced by the unchanged $I_D/I_G$ value. It further demonstrates that friction occurs at the interface of DLC/MoS$_2$, and notably, the periodic micro-topological meta-contact structures are formed at the friction interface, constituting the normalized contact interface.

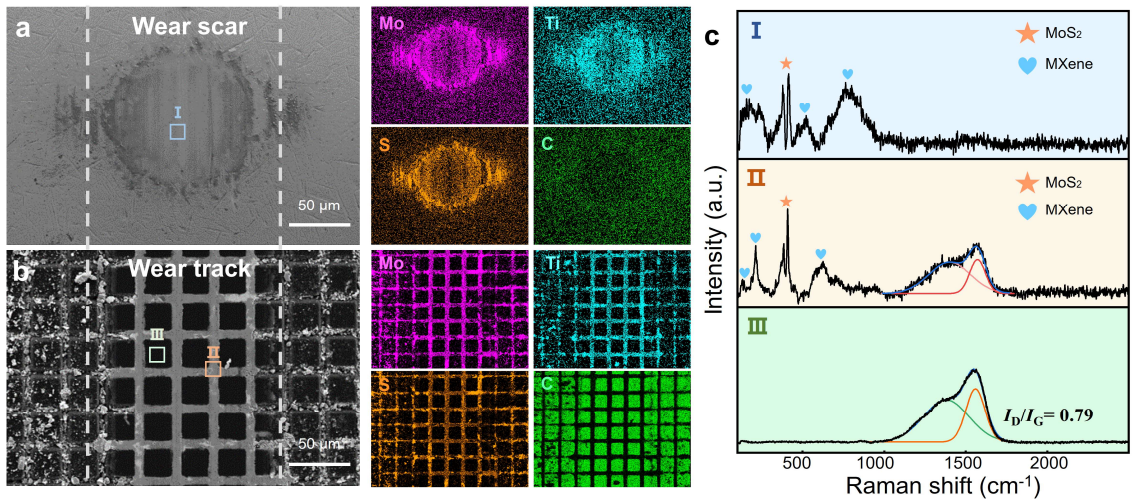

**Figure S13.** Microstructures and element mappings of friction interface. a) SEM morphologies and EDS mappings of wear scar. b) SEM morphologies and EDS mappings of wear track. c) Raman spectra of wear scar and wear track, Raman spectra of the regions marked in (a) and (b).

**8: Severe out-of-plane deformation of MoS₂ coating on the textured steel**

As shown in Figure S14, the wear track of MoS$_2$ coating on textured steel under average contact pressure of 12.7 GPa in air is severely damaged, and the steel substrate undergo deformation and is exposed.

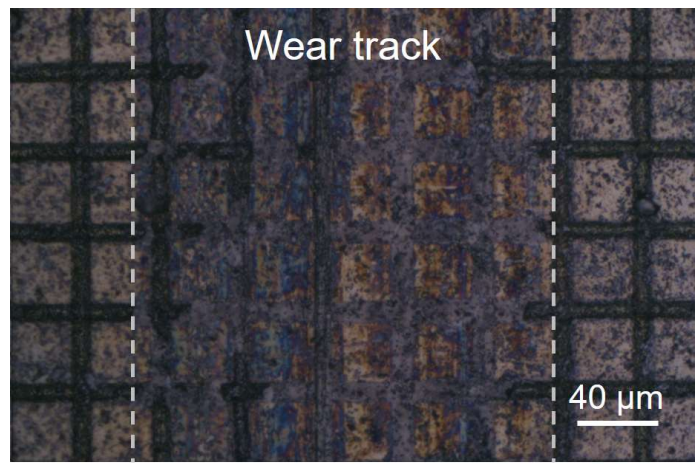

**Figure S14.** Wear track morphology of the MoS$_2$ coating on textured steel

## 9: Tribological properties of DLC/MoS$_2$/MXene composite on textured steel and DLC film

As shown in Figure S15, the DLC/MoS$_2$/MXene composite on textured steel achieves engineering-grade superlubricity at the applied loads of 15 N, 20 N, 25 N. However, as the applied load increases, the superlubricity lifespan decreases. Under an applied load of 25 N, the superlubricity persists for approximately 25,000 laps, while at 30 N, the lifespan shortens to about 15,000 laps.

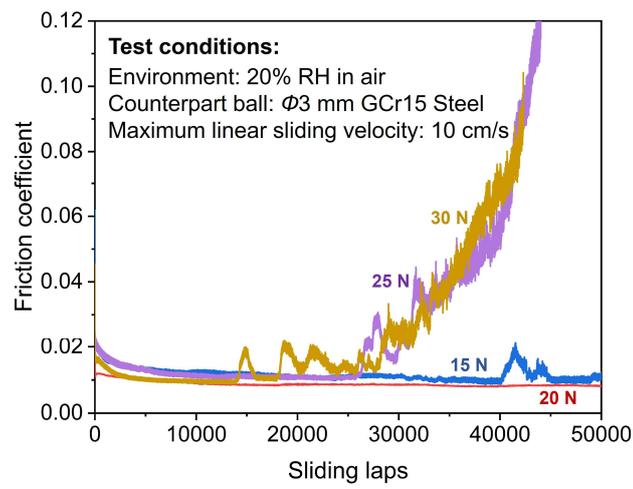

**Figure S15.** Friction curves of DLC/MoS$_2$/MXene composite on textured steel under different applied loads.

As shown in Figure S16, DLC film on steel exhibits a high friction coefficient of approximately 0.03.

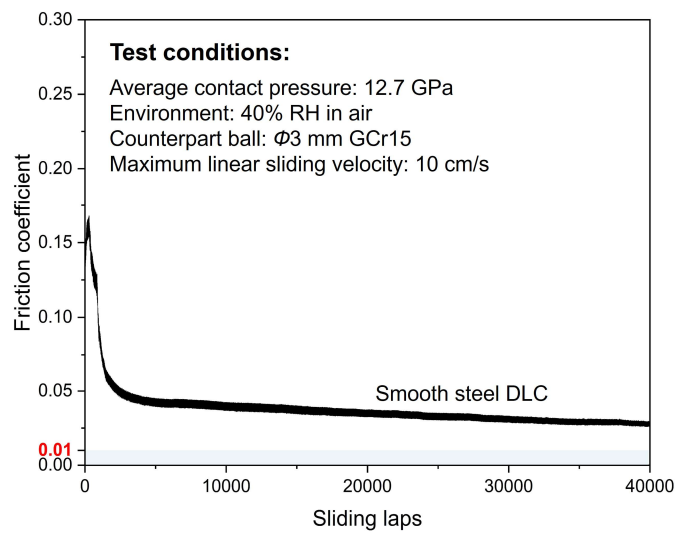

**Figure S16.** Friction curve of DLC film on steel under average contact pressure of 12.7 GPa and 40% RH in air.

Figure S17 showed the engineering-grade superlubricity of DLC/MoS$_2$/MXene composite on textured steel under a nitrogen atmosphere of 30% RH and average contact pressure of 12.7 GPa, and an ultra-long wear life of approximately 3.5×10$^5$ laps (3,500 meters) superlubricity is achieved.

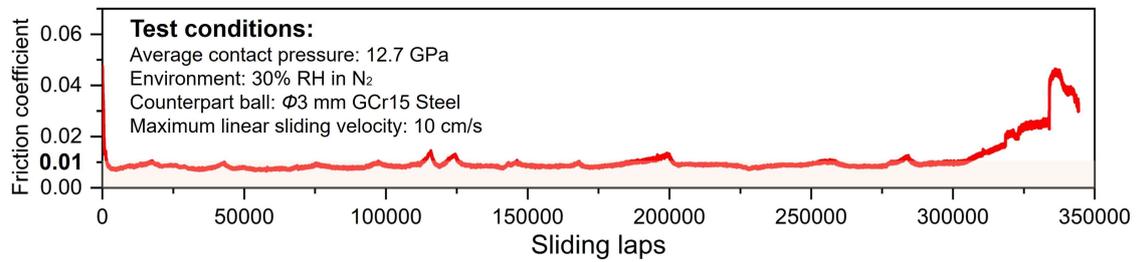

**Figure S17.** Engineering-grade superlubricity of DLC/MoS$_2$/MXene composite on textured steel.

**Captions for Movies:**

**Movie S1.**

Transition of crystalline $MoS_2/MoS_2$ contact configurations during dynamic sliding.

**Movie S2.**

Transition of amorphous DLC/crystalline $MoS_2$ contact configurations during dynamic sliding.